\documentstyle[psfig,prb,aps]{revtex}
\begin{document}
\draft

\twocolumn[\hsize\textwidth\columnwidth\hsize\csname
@twocolumnfalse\endcsname

\title{
Pseudopotential study of binding properties of solids within generalized 
gradient approximations: The role of core-valence exchange-correlation.
}
\author{M.~Fuchs, M.~Bockstedte,\cite{bockstedte} E.~Pehlke,\cite{pehlke} and
M.~Scheffler}
\address{Fritz-Haber-Institut der Max-Planck-Gesellschaft, Faradayweg 4-6,
D-14195 Berlin-Dahlem, Germany}
\date{submitted to Phys.~Rev.~B, July, 11, 1997}
\maketitle
\begin{abstract}
In {\em ab initio} pseudopotential calculations within density-functional 
theory the nonlinear exchange-correlation interaction between valence and 
core electrons is often treated linearly through the pseudopotential. 
We discuss the accuracy and limitations of this approximation regarding a 
comparison of the local density approximation (LDA) and generalized gradient 
approximations (GGA), which we find to describe core-valence exchange-correlation 
markedly different.
(1) Evaluating the binding properties of a number of typical solids we demonstrate 
that the pseudopotential approach and namely the linearization of core-valence 
exchange-correlation are both accurate and limited in the same way in GGA as in LDA. 
(2) Examining the practice to carry out GGA calculations using pseudopotentials 
derived within LDA we show that the ensuing results differ significantly from 
those obtained using pseudopotentials derived within GGA.
As principal source of these differences we identify the distinct behavior of 
core-valence exchange-correlation in LDA and GGA which, accordingly, 
contributes substantially to the GGA induced changes of calculated binding 
properties.
\end{abstract}

\vskip2pc]

\section{Introduction}

Generalized gradient approximations (GGAs) to the exchange-correlation (XC) 
energy~\cite{lan81a,per85a,bec88a,per86a,per92a} in density-functional 
theory~\cite{hoh64a,dre90a} are currently receiving growing interest as 
a simple alternative to improve over the local density approximation 
(LDA)~\cite{koh65a,jon89a} in {\em ab initio} total energy calculations. 
In various respects the GGA proved to be more appropriate than the LDA, 
namely 
(1) Binding energies of molecules~\cite{pop93a,bec92a} and 
solids~\cite{phi96a,koe92a,gar92a} turn out more accurate, 
correcting the tendency of the LDA to overbinding.
(2) Activation energy barriers, e.g. for the dissociative adsorption
of H$_2$ on metal and semiconductor surfaces~\cite{ham93a,ham94a,peh95a}
are in distinctly better accordance with experiment. 
Reaction and activation energies for a variety of chemical reactions 
show a similar improvement.~\cite{por95a,bak95a,fan92a}
(3) The relative stability of structural phases seems to be predicted more 
realistically for magnetic~\cite{leu91a} as well as for 
nonmagnetic~\cite{ham96a,mol95a} materials. 
Bulk structural properties are often not improved within GGA. 
While the lattice parameters consistently increase compared to the LDA, a 
closer agreement with experimental data is reported for alkali, $3d$, and
some $4d$ metals.~\cite{per92a,cho96a,khe95a,ozo93a} 
An overestimation of up to several percent is found however for $5d$ metals 
and common semiconductors, their bulk modulus turning out too small accordingly 
(typically by $\lesssim 25\%$).\cite{dal96a,fil94b,ort92a}

Regarding the understanding of the GGA and further advances beyond it, it 
is important to gain insight into the XC related ``mechanisms'' underlying
an eventually improved performance, e.g. along the lines 
pursued in Refs.~\onlinecite{per92a,phi96a,umr93a}, and \onlinecite{zup97b}. 
Complementary, careful estimates are needed to what degree computational 
approximations in evaluating the total energy interfere with a comparison 
of different XC functionals.
With this in mind we examine two interrelated issues that have been of 
persistent concern in comparison of the LDA and the GGA based on 
pseudopotential calculations.~\cite{pic88a}

Firstly in how far the nonlinearities associated with the XC interaction
of core and valence electrons in these XC density-functionals can be accounted 
for by the pseudopotentials. 
Computationally it is expedient to treat core-valence XC as a part of
the pseudopotential and thus as if it acted linearly on the (pseudo) valence
electron density.
Within the LDA the transferability of the pseudopotentials remains, in most
cases, intact under this approximation, i.e. good agreement of the results 
of pseudopotential with all-electron calculations may be expected without 
handling these nonlinearities explicitly.
The extent to which this carries over to GGA's, and thus enables a
menaningful comparison with the LDA, is unclear at present. 
Experience with GGA's still needs to be built up and previous studies 
have advanced conflicting views on this subject:
Examining structural parameters of crystalline solids,
Juan~{\em et al.}~\cite{jua93a,jua95a} concluded nonlinear
core corrections for XC were required in pseudopotential calculations within
the GGA by Perdew and Wang~\cite{per92a} (PW) 
even in cases where they are negligible in LDA, like bulk Si.
On the other hand, Moll~{\em et al.}~\cite{mol95a} and
Dal~Corso~{\em et al.}~\cite{dal96a} found the LDA and PW GGA
behaved alike in this respect.

Secondly we discuss the role of differences seen in the
pseudopotentials constructed within the LDA and within the GGA. 
Here we address, on the one hand, in how far such differences
are small enough to warrant the circumventing of a full 
self-consistent GGA calculation (using GGA pseudopotentials) 
by a computationally simpler {\em post}-LDA treatment 
where the electronic total energy is first minimized within LDA 
(using LDA pseudopotentials) and then corrected perturbatively
for the GGA XC energy.
On the other hand, the behavior of the pseudopotentials eventually 
reflects a different description of the core-valence interactions 
in LDA and GGA. 
This allows us to conceive the GGA-related effects separately in 
terms of XC among the valence electrons themselves and XC of the 
valence with the core electrons.
Results in several works indeed hint that the LDA and GGA might differ 
in this respect:
Garcia~{\em et al.}~\cite{gar92a} evaluated cohesive properties of
some metals and semiconductors on the basis of the Becke/Perdew
(BP) GGA~\cite{bec88a,per86a} and a precursor to the PW GGA.
Dependent on whether the pseudopotentials screened within the GGA 
were derived from an LDA or GGA calculation of the free atom they 
obtained, in some instances, differing values of the lattice parameters 
{\em and} cohesive energies.
Examining the dissociation of silanes using LDA-based pseudopotentials
Nachtigall~{\em et al.}~\cite{nac96a} observed noticeably overestimated
activation and reaction energies compared to the respective all-electron
approach for various GGAs, but close agreement for the LDA. Despite
the apparent incongruencies,
GGA calculations are still being based on LDA 
pseudopotentials.~\cite{ham96b,kra96a}

To address these issues we investigate the differences between the LDA and 
GGA systematically at each step of the pseudopotential approach, the 
construction 
of the pseudopotentials from atomic calculations and their use in polyatomic 
systems.
In turn we evaluate the cohesive properties of a set of typical
metallic, semiconducting, and insulating crystals (Na, Mg, Al, Cu, W, 
diamond, Si, Ge, GaAs, and NaCl), where we apply pseudopotentials
with and without nonlinear core corrections. 
With respect to the role of core-valence XC we establish in how far its
handling affects the accuracy of pseudopotential calculations by comparing
our results with available all-electron data. 
We then discuss the related need for the consistent use of the same XC 
scheme at all points of a pseudopotential calculation and comment on
the contribution to the GGA induced changes of LDA results for cohesive 
properties driven by differences between LDA and GGA core-valence XC.

Concerning proposals for GGAs we present results for the PW and the earlier BP 
schemes.  Both are variants of the generic type 
\begin{equation}
E^{\rm GGA}_{\rm XC}[n] = 
	\int n({\bf r}) \epsilon^{\rm GGA}_{\rm XC}(n({\bf r}),\nabla n({\bf r}))  \,
	d^3r,
\end{equation}
depending locally on the electronic density $n({\bf r})$ and its gradient 
and yielding a local XC potential 
$V_{\rm XC}({\bf r}) = \delta E_{\rm XC}[n] / \delta n({\bf r})$ 
like in case of the LDA. 
These schemes are widely used in present day applications, and remain of
interest as a starting point in recent nonlocal hybrid XC schemes 
expected to further improve over GGA type functionals.~\cite{bec96a}
Accurate all-electron results are available for those GGA schemes and serve 
as a rigorous reference for the pseudopotential calculations in this study.
The PW GGA is derived basically from first principles, combining the gradient 
expansions of the exchange and correlation holes of a non-uniform 
electron gas with real-space truncations to enforce constraints imposed 
by properties of the physical XC hole. 
While the BP GGA may be deemed to be somewhat more heuristic as it relies also 
on fitted parameters, it has been yielding results close to those of the PW 
GGA, at least in all-electron calculations. 
In addition we have considered the recently proposed GGA by Perdew, Burke and
Ernzerhof (PBE) which is regarded to be conceptually more concise than the PW 
GGA but is expected to perform essentially similarly.~\cite{per96b}
In the pseudopotential calculations for the properties addressed here we have 
found the PBE and the PW GGA to yield nearly equivalent
results,~\cite{com:pbe} hence our conclusions for the PW GGA hold for 
the PBE GGA as well. 

The remainder of this paper is organized as follows:
In Sec.~\ref{sec:theory} we briefly review and discuss the relevant formal 
aspects of pseudopotential calculations. Technical characteristics of
our calculations are outlined in Sec.~\ref{tech}. In Sec.~\ref{sec:application} 
we report our results, and put them in perspective with our considerations 
from Sec.~\ref{sec:theory}. Section~\ref{sec:summary} summarizes our
conlcusions.  Atomic units are used throughout unless indicated otherwise.

\section{General considerations}
\label{sec:theory}

In  pseudopotential calculations the total
energy is formally treated as a functional of the
valence charge density alone, with the 
pseudopotentials accounting for the interaction
of the valence electrons with the nuclei and
with the core electrons - namely for Pauli repulsion, 
electrostatic and XC interactions - to within
the frozen core approximation.~\cite{bar80a,ihm79a} 
Substituting the GGA for the LDA modifies the 
treatment not only of the XC interactions of 
the valence electrons among themselves but also
that of the core-valence (CV) interactions. In order to treat all
interactions within one and the same XC scheme the pseudopotentials
to be employed in a GGA calculation in principle
ought to be generated {\em consistently} within the same GGA 
as well, rather than within, say, the LDA. In the following
we discuss the relevance of this ``pseudopotential
consistency'' to total energy calculations in LDA and 
GGA. 
Within the pseudopotential framework the GGA total energy functional 
reads 
\begin{equation}
\label{total:energy}
E^{\rm GGA}_{\rm tot}[n]
= 
T_{0}[n] + E_{\rm H}[n] + E_{\rm XC}^{\rm GGA}[n]
+ \sum_{i}^{\rm occ} \langle \psi_{i} | \hat V^{\rm GGA} | \psi_{i} \rangle,
\end{equation}
where the various terms denote the noninteracting kinetic energy,
the Hartree energy, the XC energy and the potential energy of the 
valence electrons, represented by the pseudo 
wavefunctions $\psi_{i}({\bf r})$ and the corresponding charge
density $n({\bf r})=\sum_{i}^{\rm occ} |\psi_{i}({\bf r})|^2$ 
in the presence of the ion cores, represented by GGA pseudopotentials, $\hat 
V^{\rm GGA}$. 
The LDA counterpart to (\ref{total:energy}) is obtained
by substituting the XC energy $E_{\rm XC}^{\rm LDA}$
and the LDA pseudopotentials $\hat V^{\rm LDA}$ for the 
respective GGA entities.  
 
Now the groundstate energies in GGA and LDA can be readily 
compared with the help of a perturbative analysis of the total 
energy functionals 
$E^{\rm GGA}_{\rm tot}[n]$ 
and 
$E^{\rm LDA}_{\rm tot}[n]$,
at any given set of ionic positions. 
Around the respective 
stationary groundstates, characterized by the densities 
$n^{\rm GGA}$ and $n^{\rm LDA}$, the variational principle 
implies that
\begin{equation}
\label{stationarity}
E^{\rm GGA}_{\rm tot}[n] 
\simeq 
E^{\rm GGA}_{\rm tot}[n^{\rm GGA}] + {\cal O}[(n-n^{\rm GGA})^2],
\end{equation}
and likewise in LDA. Supposed GGA and LDA yield
similar densities the difference of their groundstate
energies, 
$\delta E_{\rm tot} = E^{\rm GGA}_{\rm tot}[n^{\rm GGA}] - E^{\rm LDA}_{\rm 
tot}[n^{\rm LDA}]$,
can be expressed by virtue of (\ref{stationarity}) as
\begin{eqnarray}
\label{1storder}
\delta E_{\rm tot} 
&\simeq&
E^{\rm GGA}_{\rm XC}[n^{\rm LDA}] - E^{\rm LDA}_{\rm XC}[n^{\rm LDA}]\\
&&
\nonumber
+ \sum_{i}^{\rm occ} \langle \psi_{i}^{\rm LDA} | \hat V^{\rm GGA} - \hat 
V^{\rm LDA} | \psi_{i}^{\rm LDA} \rangle,
\end{eqnarray}
i.e. simply in terms of the density and wavefunctions obtained
within the LDA, using the LDA pseudopotential, $\hat V^{\rm LDA}$. Accordingly 
the GGA modifies the (pseudo) 
total energy in two ways: 
(i) By the direct difference of the XC energies 
$\delta E_{\rm XC} =
      E^{\rm GGA}_{\rm XC}[n^{\rm LDA}] - E^{\rm LDA}_{\rm XC}[n^{\rm LDA}]$.
This term corresponds to the often applied {\em a posteriori} 
gradient-correction 
scheme where, at given ionic coordinates, the density is 
evaluated self-consistently from the LDA XC potential and then 
used to compute the total energy with the GGA XC energy functional. 
For the GGAs considered the $\delta E_{\rm XC}$ is negative, and
vanishes in the limiting case of the homogeneous electron gas. Typically
its magnitude increases with the degree of inhomogeneity of the 
system at hand,~\cite{mol95a} the GGA correction to the LDA XC
energy being larger for a free atom or molecule than
in a solid.
(ii) By the potential energy correction,
$\delta E_{V} =
 \sum_{i}^{\rm occ} \langle \psi_{i}^{\rm LDA} | \hat V^{\rm GGA} - \hat V^{\rm 
LDA} | \psi_{i}^{\rm LDA} \rangle$,
which arises as a consequence of the pseudopotential 
approximation and eventually reflects the differences in the behavior 
of the CV interactions in LDA and GGA. Note that in an all-electron
formulation this term would be absent altogether. 
Clearly the potential energy correction is missed when LDA pseudopotentials
are carried over to GGA calculations, giving rise to a
``portability'' error compared to the consistent GGA calculation
using GGA pseudopotentials. We shall demonstrate in 
Section~\ref{sec:application}
that $\delta E_{V}$ does not cancel out when total energy differences
are considered but is in general of similar importance as $\delta E_{\rm XC}$ 
for quantitative tests of the GGA within the pseudopotential framework. 

For an understanding of the differences between the GGA and LDA it is 
worthwile to examine more closely the various contributions to the CV 
interactions the pseudopotentials mediate.
In the following we identify and discuss these for 
norm-conserving pseudopotentials,~\cite{ham79a}
constructed by standard schemes~\cite{tro91a,bac82a} from atomic all-electron 
calculations. 
As a canonical first step 
these algorithms generate angular momentum dependent 
screened pseudopotentials $V^{\rm eff}_{l}[n_{0}]$ 
from a particular reference configuration, e.g. the 
groundstate of the neutral atom, assuming spherical 
screening. These act as effective potentials on the atomic pseudo valence states
via the radial Schr{\"o}dinger equations, 
\begin{equation}
\label{eq:radial:schroedinger}
\left(
-\frac{1}{2}\frac{d^2}{dr^2} + \frac{l(l+1)}{2r^{2}} + V^{\rm eff}_{l}[n_{0};r] 
-\varepsilon_{l} \right)
rR_{l}(r) = 0.
\end{equation}
The $V^{\rm eff}_{l}[n_{0}]$ contain a common spherical 
screening potential which is self-consistent with the {\em total} 
atomic charge density $n_{0}(r)$, comprised of the
(pseudo) valence density $n_{0}^{v}$ and
the core charge density $n^{c}_{0}$ obtained from the 
all-electron core states.
The effective potentials can be decomposed rigorously into 
the Hartree potentials $V_{\rm H}$ and the XC potential due 
to the valence and core electrons and an angular momentum 
dependent bare potential $V^{\rm bare}_{l}$ which conveys 
the nuclear attraction and the Pauli repulsion due to the core 
states; for an arbitrary valence configuration one has
\begin{eqnarray}
V^{\rm eff}_{l}[n;r]
&=&
      V^{\rm bare}_{l}(r)
            +V_{\rm H}[n^{c}_{0};r]
\nonumber
\\
&& 
            +V_{\rm H}[n^{v};r]
            +V_{\rm XC}[n^{v}\!\!+ n^{c}_{0};r],
\label{screened:potential}
\end{eqnarray}
which, in the reference configuration ($n=n^{v}_{0}+n^{c}_{0}$), 
reduces of course to the screened pseudopotentials. 
Through the nonlinearity of the XC potential in the density the effective
potential retains a dependence on the total rather than on the valence
density alone as it is ultimately a prerequisite for an efficient plane
wave representation. Customarily a further separation 
in terms of frozen core and variable valence contributions is accomplished
by ``linearizing'' the XC interaction taking  
\begin{eqnarray}
V_{\rm XC}[n^{v} \!\!+ n^{c}_{0};r] 
&\simeq& 
\nonumber
V_{\rm XC}[n^{v} + \tilde n^{c}_{0};r]\\
&&          +\left(
                  V_{\rm XC}[n^{v}_{0} \!\!+ n^{c}_{0};r]
                  -V_{\rm XC}[n^{v}_{0} \!\!+ \tilde n^{c}_{0};r]
            \right),
\label{linearized:xc:potential}
\end{eqnarray}
where the partial core density $\tilde n^{c}_{0}(r)$ serves 
as a control parameter. Choosing $n^{v}=n^{v}_{0}$ the 
screened pseudopotentials, and thus the atomic properties
in the reference configuration, are correctly recovered.
Now those terms on the RHS of (\ref{screened:potential}) 
which are independent of the valence density define the usual
pseudopotentials that are to be transferred to 
and screened according to the environment of one's target
system. Applying (\ref{linearized:xc:potential}) they 
read 
\begin{eqnarray}
V_{l}(r) 
&=&
V^{\rm bare}_{l}(r) + V_{\rm H}[n^{c}_{0};r]
\nonumber
\\ 
&&          +\left(
                  V_{\rm XC}[n^{v}_{0} \!\!+ n^{c}_{0};r]
                  -V_{\rm XC}[n^{v}_{0} \!\!+ \tilde n^{c}_{0};r]
            \right).
\label{ionic:potential}
\end{eqnarray}
The last term here comprises the core-valence XC interaction
\begin{equation}
\Delta V_{\rm XC}(r) = (V_{\rm XC}[n^{v}_{0} \!\!+ n^{c}_{0};r]
-V_{\rm XC}[n^{v}_{0} \!\!+ \tilde n^{c}_{0};r])
\label{cv:xc:potential}
\end{equation}
as represented by the ionic pseudopotential. Below we demonstrate that
it is the key quantity to understand the differences between the
ionic pseudopotentials in LDA and GGA which, in practice, are defined simply by
``unscreening'' the screened potentials according to
\begin{equation}
V_{l}(r) = V^{\rm eff}[n_{0};r]- V_{\rm H}[n^{v}_{0};r]- V_{\rm 
XC}[n^{v}_{0}\!\!+\tilde n^{c}_{0};r].
\end{equation}
with all quantities evaluated within the respective XC scheme.
Note that the transformation Eq.~(\ref{linearized:xc:potential}) 
turns the core-valence XC energy into a linear functional 
of the valence density that is absorbed in the pseudopotential contribution
to the total energy instead of being treated as a part of 
the XC energy itself. 
By experience the complete core-valence linearization, 
$\tilde n^{c}_{0}(r)=0$, has proven to be accurate for the
majority of applications within the LDA. It is expected to be 
justified for local functionals like the LDA and also the GGA if the overlap 
of the core and valence charge densities does not substantially 
change whenever chemical bonds are formed or altered.
Formally the nonlinear core-valence XC could be
regarded exactly, taking $\tilde n^{c}_{0}(r)=n^{c}_{0}(r)$,
and, correspondingly,
$E_{\rm XC} =: E_{\rm XC}[n+\tilde n^{c}_{0}]$
in the total energy functional (\ref{total:energy}),
where $\tilde n^{c}_{0}$ denotes the core charge density as
compounded from the frozen atomic core charge
densities.
Since the core states are strongly localized and sharply
peaked such a choice is beyond the realm of a plane wave
representation however.
If a complete linearization of CV XC proves insufficient, e.g. in calculations
of alkali metals~\cite{heb92a} or involving spin-polarization,~\cite{cho96a}
the nonlinearities can still be captured adequately in the chemically most
important interatomic regions with the help of a partial
core density, as was first realized by Louie {\em et al.}\cite{lou82a} 
It is tailored to coincide with the full core charge density beyond 
a suitable cutoff radius $r_{c}$ but avoids the sharply peaked 
structure close to the nucleus by a smooth cutoff function $b(r)$ 
chosen largely at computational expediency, 
{\em cf.} Ref.~\onlinecite{lou82a} and Sec.~\ref{sec:application},
\begin{equation}
\tilde n_{0}^{c}(r) 
= \left\{
\begin{array}{ll}
n_{0}^{c}(r)     & \mbox{for $r \geq r_{c}$},\\
~\\
b(r)n_{0}^{c}(r) & \mbox{for $r < r_{c}$, with $b(r) \leq 1$}.\\
\end{array}
\right.
\end{equation}
Note that the CV XC component of the ionic pseudopotentials vanishes
beyond $r_{c}$. 

\begin{figure}
\vspace{-10mm}
\psfig{figure=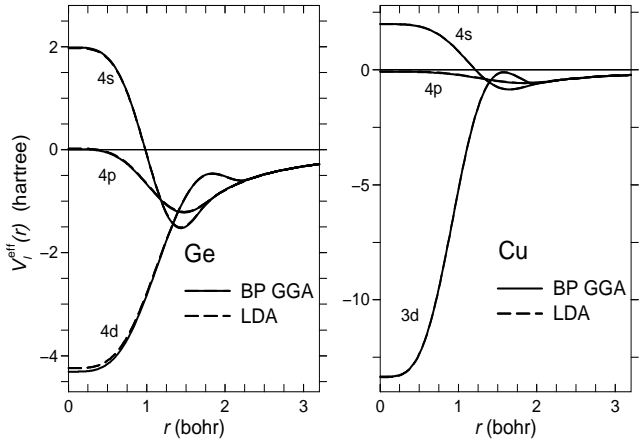,width=95mm,height=80mm,rheight=75mm}
\caption{Screened pseudopotentials within BP GGA and LDA for germanium
and copper.  On the scale of these plots the GGA and LDA pseudopotentials
lie one on top of each other.  Shown are Troullier-Martins pseudopotentials
with cutoff radii $r_{s,p}=1.9$~bohr and $r_{d}=2.3$~bohr (Ge),
and $r_{s,d}= 2.0$~bohr and $r_{p}=2.3$~bohr (Cu).
}
\label{fig:ge:screened:potential}
\end{figure}

\begin{figure}
\vspace{-10mm}
\psfig{figure=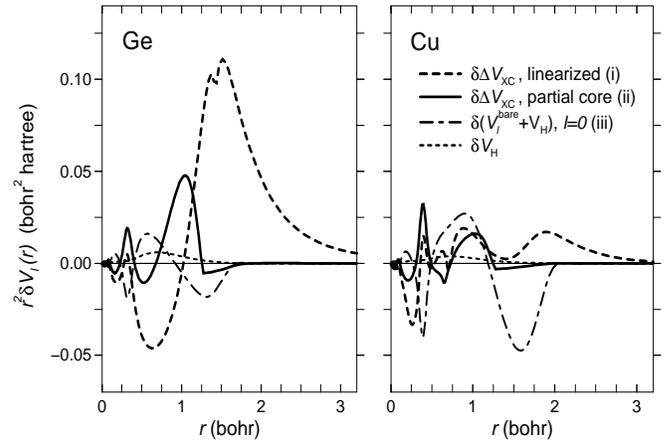,width=95mm,height=80mm,rheight=75mm}
\caption{Difference of the ionic pseudopotentials in BP GGA and LDA
for germanium and copper, corresponding to the screened potentials given in
Fig.~\protect{\ref{fig:ge:screened:potential}}. Shown are the XC, Hartree,
and bare potential contributions, {\em cf.}
Eq.~(\protect{\ref{ionic:potential}}), as discussed in the text.
Both panels use the same scale for the ordinate.
The cutoff radius of the partial core charge densities was chosen as
1.3~bohr.
}
\label{fig:ge:difference}
\end{figure}

It is well understood that the GGA does not substantially
alter the wavefunctions and the spectrum of atomic valence
states compared to the LDA, these are bound too weakly in both 
schemes,~\cite{umr93a} mainly because the XC potentials of these
schemes insufficiently cancel the repulsive contribution from
the electrons' self-interaction to the Hartree potential.
Consequently the effective potentials (\ref{screened:potential}) 
for the pseudo valence states ought to be close for both XC schemes. 
Indeed the screened LDA and GGA pseudopotentials are barely 
distinguishable by a simple visual inspection as can be seen, e.g., 
for germanium and copper in Fig.~\ref{fig:ge:screened:potential}.
%%%%%%%%%%%%%%%%%%%%%%%%%%%%%%%%%%
%
% figure 1 should appear here
%
%%%%%%%%%%%%%%%%%%%%%%%%%%%%%%%%%%
Turning to the unscreened, ionic pseudopotentials actually used for
calculations more pronounced deviations between LDA and GGA pseudopotentials
emerge.
These may by easily analyzed in terms of the various pseudopotential components
given by Eqs.~(\ref{ionic:potential}) and (\ref{linearized:xc:potential})
which contribute to the difference
$\delta V_{l}(r) = V_{l}^{\rm GGA}(r) - V_{l}^{\rm LDA}(r)$,
where the superscripts indicate the type of XC employed in constructing the 
pseudopotentials. This decomposition is illustrated in 
Fig.~\ref{fig:ge:difference},
weighting all differences with the $r$-dependent volume element.
To highlight the role of the individual contributions we distinguish three 
cases:
\begin{itemize}
\item[(i)] {completely linearized XC, $\tilde n^{c}_{0}(r)=0$.}
\item[(ii)] {approximate account of nonlinear CV XC, employing a partial 
core charge density identical with the full one outside 
$r_{c}=1.3$ bohr.}
\item[(iii)] {full account of nonlinear CV XC, taking $\tilde 
n^{c}_{0}(r)=n^{c}_{0}(r)$.}
\end{itemize}
Case (iii) serves to identify the {\em genuine} difference potential, 
due to the unlike bare and core-valence Hartree potentials. 
These originate just from the small differences of the
self-consistent atomic orbitals in LDA and GGA and thus remain
the same in (i) and (ii) where CV XC is approximated.
The core-valence Hartree potentials are seen 
to make a small positive contribution,
\[
\delta V_{\rm H}(r) = V_{\rm H}[n^{c}_{0}(\mbox{GGA});r] - V_{\rm 
H}[n^{c}_{0}(\mbox{LDA});r].
\]
%%%%%%%%%%%%%%%%%%%%%%%%%%%%%%%%%%
%
% figure 2 should appear here
%
%%%%%%%%%%%%%%%%%%%%%%%%%%%%%%%%%%
This reflects the fact that the core states are more tightly bound in GGA
than in LDA, somewhat enhancing the 
electrostatic screening of the nuclei.~\cite{kon90a,umr93a} 
Still $\delta V_{\rm H}$ is found to be weaker and faster decaying 
than the (oscillatory) differences in the bare potentials.
In case (i) the {\em genuine} difference potential is 
superimposed with a long ranged, repulsive hump.
This feature clearly signifies the distinct analytical behavior of
the CV XC potentials, Eq.~(\ref{cv:xc:potential}), in LDA and GGA, rather
than differences in the self-consistent charge densities.
Notably for germanium it attains its maximum around and stretches well 
beyond the maximum of the valence charge density up to radii that 
correspond to mid-bond positions. Including a partial core 
charge density, case (ii), eliminates by construction the core-valence 
XC potentials outside the respective cutoff radius in both LDA 
and GGA. In this region one therefore recovers the more
short ranged, and in case of germanium weaker, genuine difference potential.
Inside there remains some interference of genuine and 
core-valence XC related differences as partial and 
full core charge density deviate from each other. 

The above discussion suggests that the distinct core-valence XC 
interaction in LDA and GGA is a prime source of the differences 
between the pseudopotentials. As the LDA pseudopotentials and their GGA 
counterparts differ 
even in interatomic regions they are to be expected to perform 
unlike in a pseudopotential calculation which employs the GGA for the XC energy,
germanium with linearized CV XC being a generic example. 
By explicitly considering nonlinear CV XC, or, as exemplified by copper, by 
including
more semicore states as valence states the difference in CV XC
is removed from the pseudopotentials and instead taken into account through the
XC energy functionals themselves. In 
this case LDA and GGA pseudopotentials should thus behave more alike, provided 
of course the cooperative genuine differences are negligibly small 
themselves. In Sec.~\ref{sec:application} we substantiate these 
aspects quantitatively.

\section{Computational method}
\label{tech}

From a practitioner's point of view, GGAs are readily incorporated in
plane wave based schemes: the derivatives of the density needed to
compute XC energy and potential in position space are evaluated from
the reciprocal space representation of the density and transformed
to position space using Fourier transformations, the convergence
of all relevant quantities being controlled - like in case of LDA - through
the plane wave basis size. The construction of norm-conserving 
pseudopotentials within GGA's proceeds entirely 
parallel to the one in LDA. The necessary radial density gradients may 
be inferred, e.g., directly from the derivatives of the radial 
wavefunctions.

We have constructed the pseudopotentials~\cite{fuc97a} based on a 
scalar-relativistic 
atomic calculation using the scheme of Troullier and Martins.~\cite{tro91a}
Core and valence states were partitioned as usual, i.e. retaining
only the uppermost occupied $s$ and $p$ states as valences, except for
Cu and W where the $3d\,(5d)$ states need to be included in the valence space.
The resulting semilocal potentials were further transformed into fully 
separable representations of the Kleinman-Bylander kind.~\cite{kle82a}
In case of nonlinear CV XC we used a cuspless polynomial to represent the 
partial core charge density inside the cutoff radius. Continuity of the
density up to its third derivative is enforced to ensure that the GGA XC 
potential joins smoothly.
Various tests, carried out for the free pseudo atoms and described in the 
appendix, indicate comparable transferability for the GGA and LDA 
pseudopotentials.
The ionic pseudopotentials are tabulated and transferred without 
any intermediate fitting to the plane wave calculation.
We have refrained from any smoothing~\cite{jua93a} of the ionic GGA 
pseudopotentials
which on occasion display short-ranged oscillations, mostly 
for the PW GGA.\cite{ort92a,per96b} These correspond to a plane wave 
energy regime where the kinetic energy is dominating all other energy 
contributions and may, therefore, be conceived 
to be physically negligible. Care is required though to leave the relevant
low Fourier components intact if smoothing is performed. With a numerical 
tabulation this is attained in an unbiased, systematic manner through the
basis size cutoff.

We have computed~\cite{stu94a} the total energy per atom in the bulk systems 
varying the lattice constant within about $\pm 5\%$ of the respective
equilibrium value. 
Fitting these energies to Murnaghan's equation of state~\cite{mur44} we 
obtained the equilibrium values of the lattice parameters and the total
energy per atom. 
The cohesive energy was determined by subtracting the latter from the total 
energy of the spin-saturated spherical (pseudo) atom.
To correct this value for neglected 
contributions due to the spin-polarization of the atomic groundstate, we added
the difference of the total energies of the spin-polarized and spin-saturated 
all-electron atom within the respective XC scheme. 
Corrections of the theoretical values of the cohesive energy for the phonon 
zero-point energies are disregarded, they amount to $\approx$180~meV for 
diamond and are expected to stay below $\approx$60~meV for the other 
solids.~\cite{com:zero:phonon} 
The Brillouin zone sampling for the bulk systems was carried out using $6\times 
6\times 6$ (diamond, NaCl), $8\times 8\times 8$
(Al, Si, Ge, GaAs), and $10\times 10\times 10$ (Na, Cu, W) meshes of special 
${\bf k}$-points.~\cite{mon76a} For evaluating the cohesive energies we chose 
a plane wave cutoff energy of 50~Ry for all crystals other than diamond, Cu, 
and W for which we used 100~Ry. 
The respective structural parameters were determined with roughly two thirds of 
these values. These computational 
parameters allow for a numerical precision of better than 0.5\% for the lattice 
constants and better than 50~meV 
for the binding energies.~\cite{com:pp:scheme} 

\section{Results and discussion}
\label{sec:application}

Two sets of GGA calculations were performed where we adopted either the 
consistent GGA approach, employing the GGA for both plane wave calculation 
{\em and} construction of the the pseudopotentials, or, by contrast, the 
inconsistent
GGA approach, employing the GGA for the plane wave calculation but LDA
for the construction of the the pseudopotentials. 
The results of our calculations are compiled along with reference
data in the Tables \ref{tab:al} \-- \ref{tab:gaas}. By a comparison with
all-electron data and calculations including nonlinear CV XC we first 
demonstrate the pseudopotentials to be as transferable within the GGA
as in the LDA.
In particular transferability is not more stringently limited by 
nonlinear CV XC in GGA than in LDA. 
This provides the frame of reference for our subsequent examination of the 
consistent and inconsistent GGA approaches which are found to be inequivalent
indeed. Following our discussion of Sec.~\ref{sec:theory} we then show that
the perturbative potential energy correction yields a good quantitative account 
of the differences between these approaches. Together with the fact 
that the discrepancies are essentially eliminated once nonlinear CV XC is 
included, this allows us to identify these discrepancies as a manifestation of 
the distinct behavior of CV XC in LDA and GGA. Namely we find CV XC to 
contribute significantly to the correction of the binding energies and 
lattice parameters induced by the GGA.

\subsection{Pseudopotential transferability within GGA}

\begin{figure}
\vspace{-10mm}
\psfig{figure=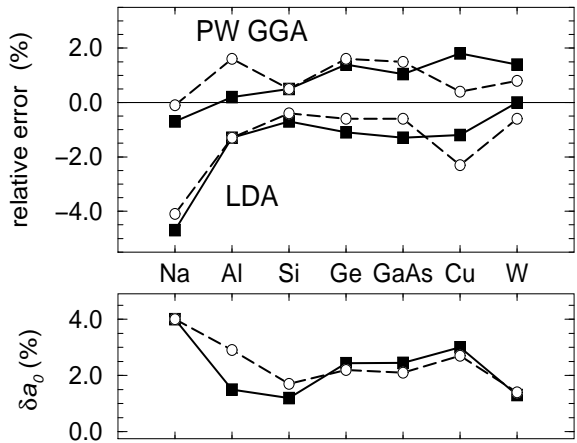,width=95mm,height=80mm,rheight=75mm}
\caption{Top panel: Relative error of the bulk lattice constant
evaluated within LDA and GGA with respect to the experimental value.
Filled squares refer to the present work, open circles
to all-electron results of Refs.~{\protect \onlinecite{khe95a} and \protect
\onlinecite{fil94b}}.
Bottom panel: Increase of the bulk lattice constant ($\delta a_{0}$)
from the LDA to the GGA value relative to the experimental value
for either method.
}
\label{fig:lapw:results}
\end{figure}

Possible uncertainties rooted in the pseudopotential approximation itself 
are properly distinguished from effects due to the use of different XC 
functionals by a comparison with all-electron data. 
To this extent we show in Fig.~\ref{fig:lapw:results} the relative error 
of the lattice constant with respect to its experimental value based on 
a compilation of results from recent all-electron 
calculations~\cite{khe95a,fil94b} and as obtained in the present 
pseudopotential framework, accounting for nonlinear CV XC and working 
with the consistent approach to PW GGA.  
It can be seen that in LDA as well as in PW GGA the results of the 
pseudopotential 
and the all-electron method agree on the order of or better than 1\%. 
In particular 
both methods yield virtually the same lattice expansion due to the GGA compared 
to LDA. Merely in case of Al the agreement is not fully quantitative.
Concerning the bulk moduli we note that the pseudopotential calculations 
reproduce the 
reduction of the LDA values due to the GGA as obtained from the all-electron 
calculations. The residual discrepancies between pseudopotential and 
all-electron results
for the bulk moduli of Ge and GaAs in LDA as well as in GGA may be seen 
as a reminder that a more accurate treatment requires the extended 3{\em d} 
states of Ga and Ge to be considered as valence states.~\cite{fil94b}
%%%%%%%%%%%%%%%%%%%%%%%%%%%%%%%%%%
%
% figure 3 should appear here
%
%%%%%%%%%%%%%%%%%%%%%%%%%%%%%%%%%%
The routine disregard of nonlinear CV XC entails no alterations of the 
calculated lattice properties compared to the results obtained with
explicit account of nonlinear CV XC for diamond, Al, and Si. For Ge
and GaAs the lattice constants
are reduced by $\approx$1\% in LDA and changed somewhat less in GGA.  
The values of the lattice constants of sodium metal turn out smaller by 
$\approx$2\% in LDA and change again less for the GGAs. Those of NaCl agree
with the experimental values at the usual level, but turn out unrealistic with
linear CV XC. 
It is well established that an explicit account 
of nonlinear CV XC is essential in order to predict dependable lattice 
properties for compounds of alkali metals within the LDA.~\cite{heb92a}
Our results for NaCl suggest that this conclusion applies to the GGAs as well.
Such a behavior seems reasonable, as in view of the relatively easily
polarizable valence shell of Na its core-valence 
overlap in metallic sodium or NaCl is likely to depart considerably from 
the one in the isolated Na atom so that an explicit account of the ensuing 
nonlinear changes of core-valence XC becomes indispensable at large. 
The presence of slight differences in the calculated lattice properties 
with and without explicit nonlinear CV XC for GaAs, Ge, and W can be similarly 
conceived as a signature of the extended core charge densities in these atoms 
compared to C, Al and Si, for which such differences are not observed.
They are likewise absent in copper, where the semicore 3{\em d}-electrons are
considered as valence states so that their XC interactions
with the 4{\em s}-electrons are incorporated exactly.
Thus we altogether find the neglect of nonlinear CV XC within the GGAs of 
similar importance to the transferability of the pseudopotentials as
within LDA.
In all cases where nonlinear CV XC is dispensable in LDA it proved to be
negligible in the GGAs as well.
Judged by the systems and properties considered we clearly 
find the pseudopotential approach itself to be equally applicable and 
accurate in GGA as in LDA.

Our results compare well with those of previous pseudopotential calculations 
reported 
in Refs.~\onlinecite{gar92a} (BP GGA) and \onlinecite{dal96a} (BP and PW GGA),
and Ref.~\onlinecite{jua95a} (PW GGA).
The severe
overcorrection of the lattice constants of Al, Si, Ge, and GaAs in case
of PW GGA without nonlinear CV XC found by Juan and Kaxiras~\cite{jua93a}
is not confirmed here. Similar to Ref.~\onlinecite{dal96a} our findings
do namely not support the conjecture put forward by these authors
in Ref.~\onlinecite{jua95a} that an explicit treatment of nonlinear CV XC were 
necessary in order to arrive at transferable pseudopotentials within 
the PW GGA. 

For the present sytems we find both GGAs to predict closely
agreeing structural properties with only immaterial differences.
Compared to the LDA we note enhanced agreement with experimental
data regarding the lattice constants of Na, Mg, Al, and Si. For
the other materials the GGA functionals overestimate the lattice constants
to a similar degree as LDA underestimates them. The bulk moduli within the
GGAs are predicted in good accordance with experiment only for Na and Cu, for
the other materials they are clearly underestimated, in particular 
(by up to $\approx$~25\%) for the semiconductors. 
In all cases the PW GGA yields cohesive energies in 
close agreement with experimental figures and corrects the 
overbinding of the LDA. Observe however that in GGA the values of the
atomic energy and hence the cohesive energy were still lowered by up
to several tenth of an eV by allowing for nonspherical groundstate
densities,~\cite{phi96a,kut87a} possibly unveilling a slight 
``underbinding'' indicated already by the overestimate of the lattice 
parameters. We note the BP GGA to yield cohesive energies 
systematically lower than in the PW GGA suggesting a slightly weaker 
binding than the PW GGA. 
  
\subsection{LDA vs. GGA pseudopotentials in GGA calculations}

Having reassured ourselves of the validity of the pseudopotential ansatz
itself we now turn to an account of the inconsistent GGA 
approach where LDA rather than GGA pseudopotentials are employed. As argued
in Section~\ref{sec:theory}, LDA and GGA pseudopotentials exhibit substantial
differences even in the interatomic regions of molecular or crystalline 
compounds. This eventually implies that LDA and GGA pseudopotentials perform 
differently 
when they are combined with the GGA XC energy functionals. 

It is evident from the results listed in Tables \ref{tab:na} - \ref{tab:gaas} 
that the outcome of the inconsistent approach depends sensitively on
the handling of core-valence XC. We shall address linear CV XC first. 
Within it, the inconsistent GGA approach yields a description of the
cohesive properties clearly disparate to that obtained from the consistent one:
the characteristic lattice expansion in either BP or PW GGA does 
not occur, and instead the lattice parameters closely resemble their LDA 
values. 
Likewise the cohesive energies turn out larger by 0.3~eV to 0.7~eV, amounting 
to $\approx$10\% (diamond) up to $\approx$50\% (Ge) of the GGA correction 
to the cohesive energy found with GGA PPs. A somewhat modified behavior is 
observed for copper where the lattice parameters are close to those from the 
consistent GGA approach and the cohesive energy correction turns out larger 
by about 0.2~eV than with GGA pseudopotentials. 
We find the consistent and the inconsistent GGA approach to both yield 
congruous 
descriptions of the binding properties once the calculations include nonlinear 
CV XC:
the inconsistent GGA approach uniformly recovers 
the typical lattice expansion as well as the decrease of the cohesive energy. 
The only incongruencies found between the two approaches concern residual
deviations of about 0.2~eV for the cohesive energies of diamond and copper.

To further discuss the GGA correction of the cohesive energy, $\delta E_{b}$, 
dependent on the choices for pseudopotential and the treatment of CV XC, we 
examine the constituent corrections of the total energies for (pseudo) atom 
and solid separately by the decomposition
\begin{eqnarray}
\delta E_{b} &\equiv& E_{b}^{\rm GGA} - E_{b}^{\rm LDA}
\\
\nonumber
&=& \delta E_{\rm tot}^{\rm atom} - \delta E_{\rm tot}^{\rm solid},
\nonumber
\end{eqnarray}
making use of the the perturbative analysis of Section \ref{sec:theory}.
Following Eq.~(\ref{1storder}) the GGA entails a twofold change of the total 
energy compared to the LDA: the direct correction of the XC energy, $\delta 
E_{\rm XC}$, 
and the potential energy correction, $\delta E_{\rm V}$.
Now the consistent approach comprises both corrections so 
that the change in total energy is given by 
$\delta E_{\rm tot} \simeq \delta E_{\rm XC} + \delta E_{\rm V}$. By contrast 
the inconsistent approach neglects the difference of the LDA and GGA 
pseudopotentials 
so that $\delta E_{\rm V}$ vanishes and the change in total energy is limited to
$\delta E_{\rm tot} \simeq \delta E_{\rm XC}$.
In Table \ref{tab:energy:analysis} we detail the various terms for some 
exemplary cases.

We first of all see the perturbative treatment to be well justified 
as it closely reproduces the total energy corrections extracted from
the respective self-consistent LDA and GGA calculations in both the 
atoms and the solids to within 0.02~eV. Adopting {\em linear} CV XC, the
XC energy correction in each case is found to account only partly for the 
GGA induced change of the cohesive energy. Instead a substantial fraction 
must be attributed to the potential energy correction and is thus indeed
effected by the difference of the LDA and GGA pseudopotentials.
Acting in a like manner as $\delta E_{\rm XC}$, $\delta E_{\rm V}$ results in
a decrease of the cohesive energy, except for copper where we observe
the above mentioned slight enlargement. 
Taking nonlinear CV XC into consideration, the magnitude of the potential 
energy 
correction is reduced compared to linear CV XC. Importantly,
$\delta E_{\rm V}$ no longer contributes to the change of the binding energy 
which is instead captured completely by the XC energy correction for Na, Si 
and Ge, shown in detail in Table~\ref{tab:energy:analysis}. It does retain
significance in case of diamond and copper however. Save for these, the
inconsistent and consistent approaches with explicit nonlinear CV XC thus 
produce alike values for the binding energies. 

We note the lattice expansion observed upon switching over from LDA to GGA 
to be consistent with the repulsive character of the difference potential
between the LDA and GGA pseudopotentials seen in the real space inspection of 
the respective pseudopotentials, {\em cf.} Fig.~\ref{fig:ge:difference}.
Similarly the potential energy correction turns out positive, and more 
so in the solid where valence charge accumulates in the bonding region.
Once nonlinear CV XC is taken into account, the XC related 
differences outside the core region are eliminated. 
What remains are essentially the genuine differences of the LDA and GGA 
pseudopotentials, reflecting the different description of the core states 
in the respective XC schemes. 
In principle the more tightly bound core states in the GGA should
make the Pauli and the Coulomb repulsion more short-ranged compared 
to the LDA, adding some repulsion about the ion sites and some attraction
at intermediate distances. The details are certainly quite species 
dependent and moreover one cannot rule out some interference from 
residual XC related differences, e.g. due to the unlike partial and 
full core densities, frustrating any {\em a priori} estimate of their
contribution to the potential energy correction $\delta E_{\rm V}$.
Nevertheless as the genuine differences are small and confined to the 
immediate vicinity of ion sites they should be rather inconsequential 
to the calculation of total energy differences. Such a scenario is conceivable 
and consistent with our results for Na, Al, Si, etc., but has its limitations 
as major portions of the valence charge density reside and adjust to
charge transfer close to the ion sites. This clearly applies for the 2{\em p} 
states of the first row elements like C and the 3{\em d} transition metals like 
Cu, 
which take on their maxima in the domain of the genuine differences. Hence
we find the potential energy correction for these elements to be only partly 
conditioned by the treatment of nonlinear CV XC.
In case of Cu the 3{\em d} electrons are considered as valence rather than
as core electrons so that their XC interactions with the 4{\em s} electrons
are accounted for exactly and not linearly through pseudopotentials. For such a 
core-valence 
partitioning it is readily verified by inspection, {\em cf.} 
Fig.~\ref{fig:ge:difference}, that the differences of LDA 
and GGA pseudopotentials are primarily of the genuine kind and thus quite 
independent of a 
further account of nonlinear CV XC with still deeper core states. Accordingly we
obtain an actually equivalent description of the bulk lattice parameters in 
GGA with either LDA or GGA pseudopotentials, with the potential energy 
correction to the 
cohesive energy being of the order of 0.2~eV within both linear and nonlinear 
CV XC. We would like to point to an analogous observation made in an 
all-electron method using pseudopotentials, where the full core density is 
retained: investigating Fe within the PW GGA, Cho {\em et al.}~\cite{cho96a} 
reported nearly identical structural and magnetic parameters with either LDA or 
GGA based pseudopotentials.

\section{Summary and Conclusions}
\label{sec:summary}

In conclusion we have shown that GGA pseudopotentials indeed convey a 
substantial share of the GGA's corrections over LDA. 
Accordingly we deem the consistent use of the GGA in the application of
the pseudopotential {\em and} their construction to be generally a key 
requirement to attain GGA quantities equivalent to those obtained within 
GGA all-electron methods. By contrast, inequivalent results arise
when the XC energy is treated in GGA but the pseudopotentials are 
taken, inconsistently, to be the same ones as in LDA. 
We have found such ``portability errors'' to be most significant when the XC
interaction of core and electrons is treated linearly as a component of the
pseudopotential, but less important when the nonlinear core-valence XC 
interaction 
is incorporated properly into the XC energy functional of the valence electrons 
itself employing a partial core density.
The precise agreement of the results for the cohesive energies from our 
perturbative
and self-consistent calculations conforms with the the common 
lore~\cite{ham93a,kon90a} 
that self-consistency has only a small effect on the value of GGA total 
energies and 
differences thereof. Instead these can be accurately evaluated with the LDA 
wavefunctions and charge density.
Our analysis shows however that such an {\em a posteriori} GGA scheme within 
the 
pseudopotential framework must treat the difference in the valence-valence
XC energies and the ionic pseudopotentials on an equal footing, as given in 
Eq.~(\ref{1storder}): 
either carrying out the initial LDA calculation with LDA pseudopotentials 
and adding to the total energy the correction
$\delta E_{\rm XC}+\delta E_{\rm V}$ 
or, equivalently, doing the LDA calculation with GGA pseudopotentials ($\delta 
E_{\rm V}\equiv 0$) and adding just $\delta E_{\rm XC}$ to the total energy.

As the differences of the pseudopotentials originate from the distinct 
core-valence XC potentials in LDA and GGA we moreover understand our findings 
as 
evidence that the bond softening in GGA is directly related to a stronger 
XC repulsion between the valence and upper core states than in the LDA. The
reduction of the binding energy by the GGA on the other hand appears to a 
larger extent due to describing the XC of the valence electrons among 
themselves within GGA instead of LDA.
The notion of a more repulsive nature of GGA core-valence XC 
agrees with and qualifies earlier observations that the GGA corrections 
to bonding properties in solids arise mainly from the immediate vicinity 
of the ions rather than from the interstitial regions.~\cite{kre94b}
Likewise it is supported by the fact that the GGA XC potential of atoms 
like the exact XC potential is superimposed with a peaked structure which 
acts repulsively at shell boundaries compared to the LDA XC 
potential.~\cite{umr93a} 
Interestingly our findings suggest that inadequacies in the description of 
core-valence XC are an important aspect of deficiencies in the description 
of chemical bonds within either LDA or GGA.

In concluding we note that a conceptual parallel of the (``inconsistent'') 
combination of LDA pseudopotentials with GGA XC is encountered in 
wavefunction-based 
many-body methods such as quantum Monte-Carlo (QMC) simulations.~\cite{qmc:rev}
In applications QMC has been mostly combined with pseudopotentials derived from 
effective one-particle schemes.~\cite{shi90a}
Thereby the interactions among the valence electrons are described exactly 
whereas the effects of the core electrons are dealt with in an approximate 
manner, say, on the level of LDA. In principle such QMC calculations provide 
an exact reference against 
which we could check the performance of approximate XC schemes like the GGA 
for the valence electrons alone. Indeed a survey of the literature indicates 
that cohesive energies for 
diamond, Si,~\cite{fah90a,li91a} and Ge~\cite{raj95a} obtained with QMC and LDA 
pseudopotentials are in significantly closer agreement with experimental 
figures than are our GGA results using LDA pseudopotentials. 
At least for these cases this
raises the question whether the GGA affords a better description of XC among 
{\em all} electrons, then yielding highly accurate binding energies, than 
of XC among the valence electrons alone. On the other hand, one is well aware 
that the use of LDA pseudopotentials in QMC simulations introduces some 
uncertainty in the QMC values for the cohesive energy 
themselves,~\cite{shi90a,raj95a} quite analogous 
to the ``portability errors'' we have found in GGA calculations using LDA 
pseudopotentials to be of the order of up to some tenths of an eV. 
On this level of accuracy it is then 
clearly desirable to obtain accurate estimates of the systematic uncertainties 
related to the use of pseudopotentials derived from DFT in exact methods 
like e.g. QMC simulations as well in order to facilitate a quantitative 
assessment of approximate XC schemes like the GGA.

\section{Appendix}
\label{sec:appendix}

In this appendix we present some tests on the transferability of our GGA 
pseudopotentials compared to the LDA ones. These serve to further
corroborate that GGA and LDA pseudopotentials show a similar inherent 
transferability
but exhibit significant differences due to core-valence (CV) XC, as discussed
in Sec.~\ref{sec:theory}. As is rather well established transferable 
pseudopotentials should closely preserve: (1) The all-electron atomic 
scattering properties 
as given by the logarithmic derivatives at some radius outside the core region 
over the range of valence energies relevant to chemical bonding, say up 
to $\pm 1$~hartree about the reference energies. (2) The all-electron atomic
hardness,~\cite{tet93a,fil95a} i.e. reproduce total energy and eigenvalues
for excited atomic configurations, to within the accuracy of the the 
underlying frozen-core approximation. 

\begin{figure}
\vspace{-10mm}
\psfig{figure=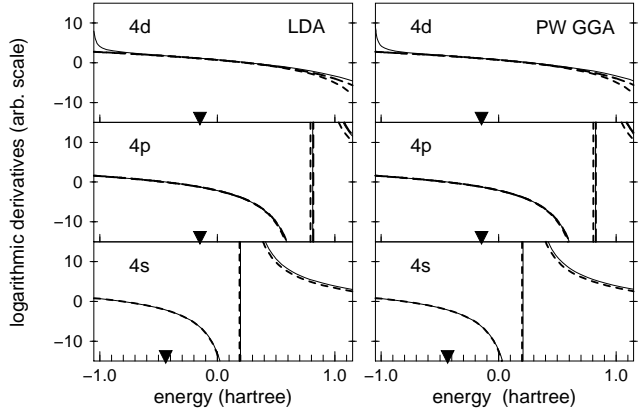,width=95mm,rheight=65mm}
\caption{Logarithmic derivatives $R_{l}'(\varepsilon)/R_{l}(\varepsilon)$
vs. energy $\varepsilon$ for the germanium atom (at $r=2.4$~bohr,
and with the $4s$-component as local potential). Solid lines correspond
to the all-electron potential, dashed lines
to the semilocal pseudopotentials, and long-dashed lines to their
Kleinman-Bylander form. Reference energies are marked by solid triangles.
In the all-electron case the pole in the $4d$-channel at $\approx -1$~hartree
is associated with the $3d$ core state.
}
\label{fig:ge:logder}
\end{figure}

In Fig.~\ref{fig:ge:logder} we show the logarithmic derivatives evaluated 
with screened pseudopotentials, {\em cf.} Eq.~(\ref{eq:radial:schroedinger}), 
taking
germanium as an example. Good 
agreement with the respective all-electron logarithmic derivatives, to be 
expected from the norm-conservation constraints, is confirmed for both LDA 
as well as GGA pseudopotentials in the semilocal and also in the 
Kleinman-Bylander 
representation. For the latter we have additionally verified the absence 
of ghost states following Gonze~{\em et al.}~\cite{gon91a} 
%%%%%%%%%%%%%%%%%%%%%%%%%%%%%%%%%%
%
% figure 4 should appear here
%
%%%%%%%%%%%%%%%%%%%%%%%%%%%%%%%%%%
%
%%%%%%%%%%%%%%%%%%%%%%%%%%%%%%%%%%
%
% figure 5 should appear here
%
%%%%%%%%%%%%%%%%%%%%%%%%%%%%%%%%%%
\begin{figure}
\vspace{-10mm}
\psfig{figure=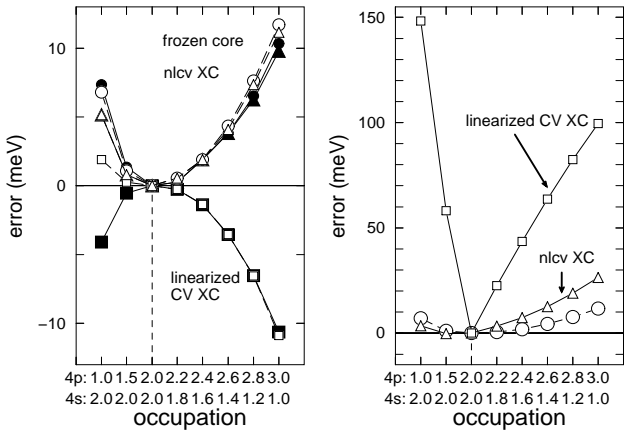,width=95mm,rheight=70mm}
\caption{
Deviations in the excitation energies of the germanium pseudo atom
compared to all-electron results, calculated from total energy differences
with respect to the groundstate configuration. The left panel corresponds to
the consistent approach using the same XC scheme throughout. The right panel
refers to the inconsistent approach, using the LDA pseudopotential but GGA for
the XC energy. Solid symbols stand for LDA, open symbols for PW GGA values.
Squares ($\Box$) correspond to calculations within linearized CV XC, and
triangles ($\triangle$) to those within nonlinear CV XC. Results obtained
within the
frozen-core approximation are shown for comparison and marked by circles
($\scriptstyle\bigcirc$). Lines are meant to guide to the eyes. Note that the
underlying
excitation energies reach up to $\approx 8$~eV.
}
\label{fig:ge:etotal}
\end{figure}
We have applied criterion (2) employing excited neutral and (positively)
ionized configuations of the spherical isolated atom.
In the case of Ge, e.g., examining
a $4s \rightarrow 4p$ electron transfer to mimic orbital hybridization
upon bond formation, and the first ionization potential. In
Fig.~\ref{fig:ge:etotal} we plot the error of the excitation energies 
for consistent calculations within LDA and GGA with respect to all-electron
calculations.

For linearized CV XC we find the ensuing errors to be of the same
magnitude albeit of opposite sign as those in a frozen-core calculation,
where only the all-electron valence states are allowed to adjust
self-consistently but the core charge density is kept fixed as that in
the atomic groundstate.
For nonlinear CV XC the errors of the pseudopotential calculation approach 
those of
the frozen-core calculation. Thus we conclude that the pseudopotential
related errors are indeed comparably small, absolutely and relatively, as
those due to neglect of core relaxation, the excitation
energy for the transfer $(4s^2,4p^2) \rightarrow (4s^1,4p^3)$ being about
$8$~eV.
Carrying out the tests in inconsistent manner -- using GGA XC and LDA 
pseudopotentials
with linearized CV XC -- leads to large devations.
%%%%%%%%%%%%%%%%%%%%%%%%%%%%%%%%%%
%
% figure 6 should appear here
%
%%%%%%%%%%%%%%%%%%%%%%%%%%%%%%%%%%
\begin{figure}
\vspace{-10mm}
\psfig{figure=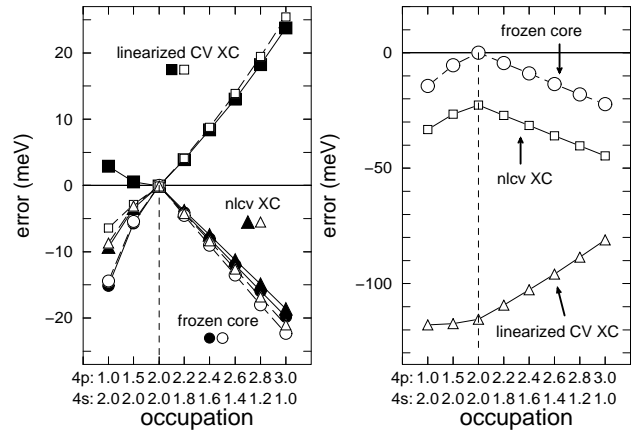,width=95mm,rheight=70mm}
\caption{Deviations of the level spacing of the $4s$ and $4p$ states of
the germanium atom with respect to all-electron results. See Fig.~
\protect{\ref{fig:ge:etotal}} for a legend.
The level spacing is $\approx 8$~eV
for the groundstate configuration, and varies by $\approx 0.8$~eV.}
\label{fig:ge:elevel}
\end{figure}
Particularly excitation energies are overestimated compared to full GGA
calculations, suggesting that the pseudopotentials in LDA are more attractive 
than in GGA, in accordance with our findings in Sec.~\ref{sec:theory} and 
\ref{sec:application}. Employing
LDA pseudopotentials together with nonlinear CV XC we obtain agreement with the 
GGA frozen-core calculation however.
Considering the eigenvalues we have observed an analogous pattern.
This is illustrated in Fig.~\ref{fig:ge:elevel} for the deviation of the 
level spacing 
$\varepsilon{4p}-\varepsilon{4s}$ with respect to all-electron calculations 
within LDA and GGA. Again a large error occurs in case of the inconsistent 
GGA calulation using LDA pseudopotentials and linearized CV XC, while the 
consistent 
approach is accurate, errors being of the order of a few ten meV compared
to changes in the level spacing of about 0.8~eV.

In summary these tests affirm our conclusion reached for the bulk systems: 
Within GGA and LDA the respective pseudopotentials possess similar 
transferability,
errors due to the usual linearization of CV XC being small; LDA 
pseudopotentials are 
adequate in GGA calculations (and vice versa) only if CV XC, acting more
repulsively in GGA compared to LDA, is incorporated explicitly.

%------- begin tabular material --------

\begin{table}
\caption{Cohesive properties of Na. 
The first column indicates the XC scheme used to generate the
pseudopotentials, the second the one employed for the XC energy 
of the (pseudo) atom and solid. Bracketed values are based on nonlinear 
core-valence XC.
We show the lattice constant $a_0$, the bulk modulus $B_0$, and the
cohesive energy $E_b$. The latter includes a spin correction of the free
Na atom of 0.20~eV (LDA) and 0.22~eV (GGA).}
\label{tab:na}
\begin{tabular}{llccc}
potential & $E_{\rm XC}$ & $a_0$ (\AA) & $B_0$ (GPa) & $E_b$ (eV) \\
\hline
LDA   & LDA &  3.98 (4.05) & 8.7 (9.1)  & 1.28 (1.22)\\
LDA   & BP  &  3.97 (4.22) & 8.7 (7.3)  & 1.06 (0.94)\\
BP    & BP  &  4.20 (4.22) & 7.3 (7.4)  & 0.94 (0.94)\\
LDA   & PW  &  3.98 (4.21) & 8.7 (7.3)  & 1.19 (1.08)\\
PW    & PW  &  4.24 (4.21) & 7.0 (7.3)  & 1.05 (1.08)\\
\hline
\multicolumn{2}{c}{LDA\tablenotemark[1]}
            &4.05        & 9.2    \\
\multicolumn{2}{c}{PW\tablenotemark[1]}
            & 4.22        & 7.1    \\
\hline
\multicolumn{2}{c}{Experiment\tablenotemark[2]}
            &  4.23       & 6.92  & 1.11\\
\end{tabular}
\tablenotetext[1]{All-electron data from Reference~\onlinecite{per92a}.}
\tablenotetext[2]{Reference~\onlinecite{com:data}.}
\end{table}

\begin{table}
\caption{Cohesive properties of NaCl. Like Table~\protect{\ref{tab:na}},
and using a spin correction of 0.20~eV (LDA) and 0.22~eV (GGA) for the
free Cl atom.}
\label{tab:nacl}
\begin{tabular}{llccc}
potential & $E_{\rm XC}$ & $a_0$ (\AA) & $B_0$ (GPa) & $E_b$ (eV) \\
\hline
LDA   & LDA & 5.19 (5.43) & 32 (32)  & 7.28 (6.99)\\
LDA   & BP  & 5.27 (5.67) & 27 (22)  & 6.59 (6.24)\\
BP    & BP  & 5.74 (5.68) & 21 (22)  & 6.20 (6.19)\\
LDA   & PW  & 5.28 (5.65) & 28 (23)  & 6.85 (6.43)\\
PW    & PW  & 5.87 (5.66) & 18 (23)  & 6.22 (6.40)\\
\hline
\multicolumn{2}{c}{Experiment\tablenotemark[1]}
      &  5.64 & 24.5  & 6.51\\
\end{tabular}
\tablenotetext[1]{Reference~\onlinecite{com:data}.}
\end{table}

\begin{table}
\caption{Cohesive properties of hcp Mg. Like Table~{\protect \ref{tab:na}}.
The equilibrium $c/a$ ratio was obtained as 1.59 (LDA) and 1.66 (GGA), the
experimental value is 1.62.{\protect \tablenotemark[1]}
}
\label{tab:mg}
\begin{tabular}{llccc}
potential & $E_{\rm XC}$ & $a_0$ (\AA) & $B_0$ (GPa) & $E_b$ (eV) \\
\hline
LDA   & LDA &  3.05 (3.16) & 39 (37)  & 2.09 (1.76)\\
BP    & BP  &  3.17 (3.18) & 32 (31)  & 1.27 (1.22)\\
PW    & PW  &  3.20 (3.20) & 30 (32)  & 1.42 (1.40)\\
\hline
\multicolumn{2}{c}{Experiment\tablenotemark[1]}
      &  3.21        & 35.4     & 1.51\\
\end{tabular}
\tablenotetext[1]{Reference~\onlinecite{com:data}.}
\end{table}

\begin{table}
\caption{Cohesive properties of Al. Like Table~{\protect \ref{tab:na}},
including a spin correction of the free Al atom of 0.15~eV (LDA) and 
0.19~eV (GGA).}
\label{tab:al}
\begin{tabular}{llccc}
potential & $E_{\rm XC}$ & $a_0$ (\AA) & $B_0$ (GPa) & $E_b$ (eV) \\ 
\hline
LDA	& LDA	& 3.97 (3.97) & 83 (85) & 4.09 (4.09)\\
LDA   & BP  & 3.97 (4.03) & 80 (75) & 3.39 (3.27)\\   
BP 	& BP	& 4.05 (4.05) & 75 (75) & 3.26 (3.25)\\
LDA   & PW  & 3.97 (4.02) & 81 (77) & 3.64 (3.54)\\   
PW	& PW	& 4.05 (4.04) & 79 (79) & 3.52 (3.53)\\
\hline
\multicolumn{2}{c}{LDA\tablenotemark[1]}
            & 3.98        & 83.9\\
\multicolumn{2}{c}{PW\tablenotemark[1]}
            & 4.10        & 72.6\\
\hline
\multicolumn{2}{c}{Experiment\tablenotemark[2]} 
            & 4.05        & 77.3	& 3.39\
\end{tabular}
\tablenotetext[1]{All-electron data from Reference \onlinecite{khe95a}.}
\tablenotetext[2]{Reference~\onlinecite{com:data}.}
\end{table}

\begin{table}
\caption{Cohesive properties of diamond. Like Table~{\protect \ref{tab:na}},
using a spin correction of 1.13~eV (LDA) and 1.26~eV (GGA) for the C atom.} 
\label{tab:diamond}
\begin{tabular}{llccc}
potential & $E_{\rm XC}$ & $a_0$ (\AA) & $B_0$ (GPa) & $E_b$ (eV) \\
\hline
LDA & LDA & 3.54 (3.54) & 436 (436) & 8.96 (8.93)  \\
LDA & BP  & 3.55 (3.58) & 421 (406) & 7.93 (7.57)  \\ 
BP  & BP  & 3.59 (3.59) & 399 (400) & 7.56 (7.58)  \\
LDA & PW  & 3.54 (3.58) & 424 (406) & 8.09 (7.92)  \\
PW  & PW  & 3.58 (3.58) & 408 (405) & 7.80 (7.83)  \\
\hline
\multicolumn{2}{c}{Experiment\tablenotemark[1]}
          & 3.57 & 442 &  7.37 \\
\end{tabular}
\tablenotetext[1]{Reference~\onlinecite{com:data}.}
\end{table}

\begin{table}
\caption{Cohesive properties of Si. Like Table~{\protect \ref{tab:na}},
using a spin correction of 0.66~eV (LDA) and 0.79~eV (GGA) for the Si atom.} 
\label{tab:si}
\begin{tabular}{llccc}
potential & $E_{\rm XC}$ & $a_0$ (\AA) & $B_0$ (GPa) & $E_b$ (eV) \\
\hline
LDA & LDA & 5.38 (5.39) & 94 (94) & 5.34 (5.32) \\
LDA & BP  & 5.40 (5.46) & 91 (86) & 4.60 (4.47) \\       
BP  & BP  & 5.47 (5.47) & 85 (85) & 4.46 (4.45) \\
LDA & PW  & 5.39 (5.45) & 92 (87) & 4.79 (4.66) \\
PW  & PW  & 5.46 (5.46) & 87 (87) & 4.64 (4.64) \\
\hline
\multicolumn{2}{c}{LDA\tablenotemark[1]}
            & 5.41        & 96   & 5.28 \\
\multicolumn{2}{c}{BP\tablenotemark[1]}
            & 5.54        & 80   \\
\multicolumn{2}{c}{PW\tablenotemark[1]}
            & 5.50        & 83   \\
\hline
\multicolumn{2}{c}{Experiment\tablenotemark[2]}
            & 5.43 & 98.8    & 4.63\\
\end{tabular}
\tablenotetext[1]{All-electron data from Reference~\onlinecite{fil94b}.}
\tablenotetext[2]{Reference~\onlinecite{com:data}.}
\end{table}

\begin{table}
\caption{Cohesive properties of Ge.
Like Table~{\protect \ref{tab:na}},
using a spin correction of 0.60~eV (LDA) and 0.74~eV (GGA) for the
Ge atom}
\label{tab:ge}
\begin{tabular}{llccc}
potential & $E_{\rm XC}$ & $a_0$ (\AA) & $B_0$ (GPa) & $E_b$ (eV) \\
\hline
LDA & LDA & 5.56 (5.60) &  73 (71) & 4.75 (4.58) \\
LDA & BP  & 5.59 (5.74) &  67 (57) & 3.96 (3.66) \\
BP  & BP  & 5.73 (5.76) &  59 (56) & 3.70 (3.66) \\
LDA & PW  & 5.58 (5.74) &  69 (59) & 4.14 (3.82) \\
PW  & PW  & 5.74 (5.74) &  58 (58) & 3.82 (3.82) \\
\hline
\multicolumn{2}{c}{LDA\tablenotemark[1]}
            & 5.63        & 78     & 4.54\\
\multicolumn{2}{c}{BP\tablenotemark[1]}
            & 5.76        & 60   \\
\multicolumn{2}{c}{PW\tablenotemark[1]}
            & 5.75        & 61   \\
\hline
\multicolumn{2}{c}{Experiment\tablenotemark[2]}
            & 5.66 &  76.8  & 3.85\\
\end{tabular}
\tablenotetext[1]{All-electron data from Reference \onlinecite{fil94b}.}
\tablenotetext[2]{Reference~\onlinecite{com:data}.}
\end{table}

\begin{table}
\caption{Cohesive properties of GaAs.
Like Table~{\protect \ref{tab:na}},
using a spin correction of 0.15~eV (LDA) and 0.18~eV (GGA) for the
Ga, and 1.41~eV (LDA) and 1.67~eV (GGA) for the As atom.}
\label{tab:gaas}
\begin{tabular}{llccc}
potential & $E_{\rm XC}$ & $a_0$ (\AA) & $B_0$ (GPa) & $E_b$ (eV) \\
\hline % troullier-martins
LDA & LDA & 5.50 (5.57) &  79 (75) & 8.68 (8.15) \\
LDA & BP  & 5.52 (5.72) &  74 (62) & 7.19 (6.33) \\
BP  & BP  & 5.68 (5.72) &  64 (62) & 6.52 (6.33) \\
LDA & PW  & 5.51 (5.70) &  76 (64) & 7.51 (6.63) \\
PW  & PW  & 5.69 (5.71) &  63 (64) & 6.74 (6.63) \\
\hline
\multicolumn{2}{c}{LDA\tablenotemark[1]}
            & 5.62        & 74     & 7.99\\
\multicolumn{2}{c}{BP\tablenotemark[1]}
            & 5.76        & 60   \\
\multicolumn{2}{c}{PW\tablenotemark[1]}
            & 5.74        & 65   \\
\hline
\multicolumn{2}{c}{Experiment\tablenotemark[2]}
            & 5.65        &  74.8    & 6.52\\
\end{tabular}
\tablenotetext[1]{All-electron data from Reference~\onlinecite{fil94b}.}
\tablenotetext[2]{Reference~\onlinecite{com:data}.}
\end{table}

\begin{table}
\caption{Cohesive properties of fcc Cu.
Like Table~{\protect \ref{tab:na}}, using a spin correction of 
0.20~eV (LDA) and 0.25~eV (GGA).
}
\label{tab:cu}
\begin{tabular}{llccc}
potential & $E_{\rm XC}$ & $a_0$ (\AA) & $B_0$ (GPa) & $E_b$ (eV) \\
\hline
LDA & LDA  & 3.55 (3.56) & 172 (172) & 4.31 (4.24) \\
LDA & BP   & 3.68 (3.70) & 124 (122) & 3.09 (3.06) \\
BP  & BP   & 3.67 (3.68) & 130 (131) & 3.22 (3.23) \\
LDA  & PW  & 3.67 (3.69) & 127 (123) & 3.23 (3.20) \\
PW  & PW   & 3.67 (3.67) & 134 (132) & 3.38 (3.38) \\
\hline
\multicolumn{2}{c}{LDA\tablenotemark[1]}
            & 3.52        & 192 & 4.29 \\
\multicolumn{2}{c}{BP\tablenotemark[1]}
            & 3.62        & 151 & 3.12 \\
\multicolumn{2}{c}{PW\tablenotemark[1]}
            & 3.62        & 151 & 3.30 \\
\hline 
\multicolumn{2}{c}{Experiment\tablenotemark[2]}
            & 3.60        & 138 & 3.50\\
\end{tabular}
\tablenotetext[1]{
All-electron values from Ref.~\onlinecite{khe95a} ($a_{0}$,
$B_{0}$), and Ref.~\onlinecite{phi96a} ($E_b$).}
\tablenotetext[2]{Reference~\onlinecite{com:data}.}
\end{table}

\begin{table}
\caption{Cohesive properties of bcc W. 
Like Table~{\protect \ref{tab:na}}, using a spin correction of
2.04~eV (LDA) and 2.32~eV (GGA).
}
\label{tab:w}
\begin{tabular}{llccc}
potential & $E_{\rm XC}$ & $a_0$ (\AA) & $B_0$ (GPa) & $E_b$ (eV) \\
\hline
LDA & LDA  & 3.14 (3.16) & 324 (331) & 10.76 (10.24)\\
LDA & BP   & 3.15 (3.21) & 308 (306) &  9.26 (8.52) \\
BP  & BP   & 3.20 (3.21) & 299 (308) &  8.73 (8.54) \\
LDA & PW   & 3.14 (3.20) & 313 (308) &  9.63 (8.87) \\
PW  & PW   & 3.22 (3.21) & 298 (310) &  8.88 (8.87) \\
\hline
\multicolumn{2}{c}{LDA\tablenotemark[1]}
           & 3.14 & 337 \\%khe95a
\multicolumn{2}{c}{PW\tablenotemark[1]}
           & 3.19 & 307 \\%khe95a
\hline
\multicolumn{2}{c}{Experiment\tablenotemark[2]}
           & 3.16 & 310 & 8.90 \\ %khe95a khe95a kittel
\end{tabular}
\tablenotetext[1]{All-electron values from Ref.~\onlinecite{khe95a}.}
\tablenotetext[2]{Reference~\onlinecite{com:data}.}
\end{table}

\widetext
\begin{table}
\caption{Change of the cohesive energy, $\delta E_b$, 
due to the replacement of the LDA by the PW GGA.
We list the total energy change per atom as obtained from self-consistent 
calculations, $\delta E_{\rm tot}$, and according to 
Eq.~(\protect{\ref{1storder}}),
$\delta E_{\rm tot}\simeq \delta E_{\rm XC}  + \delta E_{\rm V}$. 
The latter is decomposed into
its consituent terms arising from the different XC energies,
$\delta E_{\rm XC}$, 
and from the different pseudopotentials, $\delta E_{\rm V}$.  
The values without and with brackets correspond to calculations 
with linearized core-valence XC and nonlinear core-valence XC respectively.
No spin-corrections were applied to the atomic energies. The lattice 
constants were kept at their experimental values in all calculations, the 
ensuing error of the total energy due to the deviation from the theoretical 
equilibrium structure staying below $\approx$50~meV per atom. 
Symbols are explained further in the text.}
\label{tab:energy:analysis}
\begin{tabular}{lcddd}
&     & $\delta E^{\rm solid}_{\rm tot}$ (eV) 
& $\delta E^{\rm atom}_{\rm tot}$ (eV) & $\delta E_b$ (eV)  \\
\hline
Na&$\delta E_{\rm tot}$  & 0.32 (-3.21)  & 0.10 (-3.35) & -0.22 (-0.14) \\
\\
&$\delta E_{\rm XC} + \delta E_{\rm V}$ 
& 0.32 (-3.20) &  0.10 (-3.35 ) & -0.22 (-0.15) \\
&$\delta E_{\rm XC}$  
& -0.01 (-3.20) & -0.07 (-3.34)   & -0.06 (-0.14) \\
&$\delta E_{\rm V}$   
&  0.33 (-0.00) &  0.17 ( 0.01)   & -0.16 ( 0.01)\\ 
%
%Na&$\delta E_{\rm tot}$ & 0.32028 -3.207621 & 0.96386  -3.35244
%&$\delta E_{\rm XC}$ & -0.0064 -3.19856 & -.0696 -3.341212
%&$\delta E_{\rm V}$  &  0.3275 -0.00026  &.169093 -.008227
%
\hline
Diamond &$\delta E_{\rm tot}$ & 0.31 (-1.42) & -0.71 (-2.42)  & -1.02 (-1.00) \\
\\
&$\delta E_{\rm XC} + \delta E_{\rm V}$ 
& 0.32 (-1.40) & -0.70 (-2.41 ) & -1.02 (-1.01) \\
&$\delta E_{\rm XC}$  
& -0.46 (-1.37) & -1.19 (-1.34)   & -0.74 (-0.81) \\
&$\delta E_{\rm V}$   
&  0.78 (-0.03) &  0.49 (-0.13)   & -0.29 (-0.10) \\ 
\hline
Si&$\delta E_{\rm tot}$ & 0.60 (-0.97) & 0.03 (-1.52)  & -0.57 (-0.55) \\
\\
&$\delta E_{\rm XC}  + \delta E_{\rm V}$ 
& 0.61 (-0.96) & 0.04 (-1.51) & -0.57 (-0.55)\\
&$\delta E_{\rm XC}$ 
& -0.27 (-1.09) & -0.69 (-1.62) & -0.42 (-0.53)\\
&$\delta E_{\rm V}$  
&  0.88 ( 0.13) &  0.73 ( 0.11) & -0.15 (-0.02)\\
\hline
Ge     &$\delta E_{\rm tot}$ & 1.50 (-5.50) & 0.65 (-6.16) & -0.85 (-0.66) \\
\\
&$\delta E_{\rm XC} + \delta E_{\rm V}$ 
& 1.50 (-5.48) & 0.65 (-6.16) & -0.85 (-0.68) \\
&$\delta E_{\rm XC}$ 
& -0.24 (-5.49) & -0.66 (-6.17)& -0.42 (-0.68) \\
&$\delta E_{\rm V}$  
&  1.74 ( 0.01) &  1.31 ( 0.01)& -0.43 ( 0.00) \\
\hline
Cu     &$\delta E_{\rm tot}$ & 1.63 (-1.99) & 0.69 (-2.96) & -0.94 (-0.97) \\
\\
&$\delta E_{\rm XC} + \delta E_{\rm V}$ 
&  1.64 (-1.99) & 0.70 (-2.95) & -0.95 (-0.96) \\
&$\delta E_{\rm XC}$ 
&  -5.54 (-7.48) & -6.66 (-8.64) & -1.12 (-1.16) \\
&$\delta E_{\rm V}$  
&   7.19 ( 5.50) &  7.36 ( 5.69) &  0.17 ( 0.19) \\
\end{tabular}
\end{table}  
\narrowtext

%------- end tabular material ----------

\end{document}